\input psfig
\def\mev{\,{\rm Me\kern-0.1em V}}
\def\gev{\,{\rm Ge\kern-0.1em V}}

\documentstyle[11pt]{article}

\begin{document}
\vspace*{-1.25in}
\small{
\begin{flushright}
FERMILAB-PUB-96/016-T \\[-.1in] 
January, 1996 \\
\end{flushright}}
\vspace*{.85in}
\begin{center}
{\Large{\bf Electromagnetic Splittings and Light Quark Masses \\
 in Lattice QCD \\}
}
\vspace*{.45in}
{\large{A.~Duncan$^1$, E.~Eichten$^2$ and H.~Thacker$^3$}} \\ 
\vspace*{.15in}
$^1$Dept. of Physics and Astronomy, Univ. of Pittsburgh, Pittsburgh, PA 15620\\
$^2$Fermilab, P.O. Box 500, Batavia, IL 60510 \\
$^3$Dept.of Physics, University of Virginia, Charlottesville, VA 22901
\end{center}
\vspace*{.3in}
\begin{abstract}
A method for computing electromagnetic properties of hadrons 
in lattice QCD is described and preliminary numerical results are
presented. The electromagnetic field is introduced dynamically, using
a noncompact formulation.  Employing enhanced electric charges, 
the dependence of the pseudoscalar meson mass on 
the (anti)quark charges and masses can be accurately calculated.
At $\beta=5.7$ with Wilson action,
the $\pi^+-\pi^0$ splitting is found to be $4.9(3)$ MeV. 
Using the measured $K^0-K^+$ splitting, we also find 
$m_u/m_d = .512(6)$.  Systematic errors are discussed.
\end{abstract}


\newpage


If a fundamental theory of quark masses ever emerges, it may be as important
to resolve the theoretical uncertainty in the light quark masses as it is to accurately
measure the top quark mass. Moreover, an accurate determination of the up quark
mass might finally resolve the question of whether nature avoids the strong CP 
problem via a massless up quark. The particle data tables \cite{pdata}
give wide ranges for the up ($2 < m_u <8$ MeV) and down ($5 < m_d <15$ MeV) quarks,
while lowest order chiral perturbation theory 
\cite{Weinberg,leutrev,Donoghue} gives $m_u/m_d=0.57\pm 0.04$.
Numerical lattice calculations provide, in principle, 
a very precise way of studying the dependence
of hadron masses on the lagrangian quark mass parameters\cite{quark_mass}. 
However, the contribution to hadronic
mass splittings within isomultiplets from electromagnetic 
(virtual photon) effects 
is comparable to the size of the up-down quark  mass difference. 
Thus an accurate determination of the light quark masses 
requires the calculation of electromagnetic effects in the context of
nonperturbative QCD dynamics. In this letter, we discuss a 
method for studying electromagnetic effects in the hadron spectrum. In addition
to the SU(3) color gauge field, we introduce a U(1) electromagnetic field on the
lattice which is also treated by Monte Carlo methods. The resulting SU(3)$\times$U(1)
gauge configurations are then analyzed by standard hadron propagator techniques. 
  
The small size of electromagnetic mass splittings makes their
accurate determination by conventional lattice techniques 
difficult if the electromagnetic coupling is taken at its physical value. 
One of the main results of this paper is to demonstrate that calculations
done at larger values of the quark electric charges (roughly 2 to 6 times
physical values)
lead to accurately measurable electromagnetic splittings in the light pseudoscalar meson
spectrum, while still allowing  perturbative extrapolation to  physical values. 

The strategy of the calculation is as follows.  Quark 
propagators are generated in the
presence of background SU(3)$\times$U(1) fields where 
the SU(3) component represents the usual gluonic
gauge degrees of freedom, while the U(1) component incorporates an abelian
photon field (with a noncompact gauge action) 
which interacts with quarks of specified electric charge. 
All calculations are performed in the quenched approximation and  
Coulomb gauge is used throughout for both components. 
Quark propagators are calculated for a variety of electric charges
and light quark mass values. 
The gauge configurations were generated at $\beta=5.7$ on a
$12^3\times 24$ lattice. 200 configurations each separated by 1000 Monte
Carlo sweeps were used. In the results reported here, 
we have used four different values of charge 
given by $e_q =$0, -0.4, +0.8, and -1.2 in units in which the electron charge
is $e=\sqrt{4\pi/137} =.3028\ldots\;$. 
For each quark charge we calculate propagators for
three light quark mass values in order to allow a chiral extrapolation. 
From the resulting 12 quark propagators, 144 
quark-antiquark combinations can be formed. 
The meson propagators are then computed and masses for the 78 independent
states extracted. 

Once the full set of meson masses is computed, the analysis proceeds by a combination
of chiral and QED perturbation theory. 
In pure QCD it is known  that, in the range of masses considered here, 
the square of the pseudoscalar
meson mass is quite accurately fit by a linear function 
of the bare quark masses\cite{Xlogs}. 
We have found that this linearity in
the bare quark mass persists even in the presence of electromagnetism. For each of
the charge combinations studied, the dependence of the squared meson mass on the bare
quark mass is well described by lowest order chiral perturbation theory. Thus we 
write the pseudoscalar mass squared as
\begin{equation}
\label{eq:ChPT}
m_{P}^2 = A(e_q, e_{\bar{q}}) +  m_qB(e_q,e_{\bar{q}}) +m_{\bar{q}}B(e_{\bar{q}},e_q)
\end{equation}
where $e_q, e_{\bar{q}}$ are the quark and antiquark charges, and $m_q, m_{\bar{q}}$
are the bare quark masses, defined in terms of the Wilson hopping parameter by
$(\kappa^{-1} - \kappa^{-1}_c)/2a$. (Here $a$ is the lattice spacing.)
Because of the electromagnetic self-energy shift, 
the value of the critical hopping parameter must be determined
independently for each quark charge. This is done by requiring that the mass of
the neutral pseudoscalar meson vanish at $\kappa=\kappa_c$, as discussed below.
The results for the neutral pseudoscalars are shown in Figure 1.
\begin{figure}
$$\psfig{figure=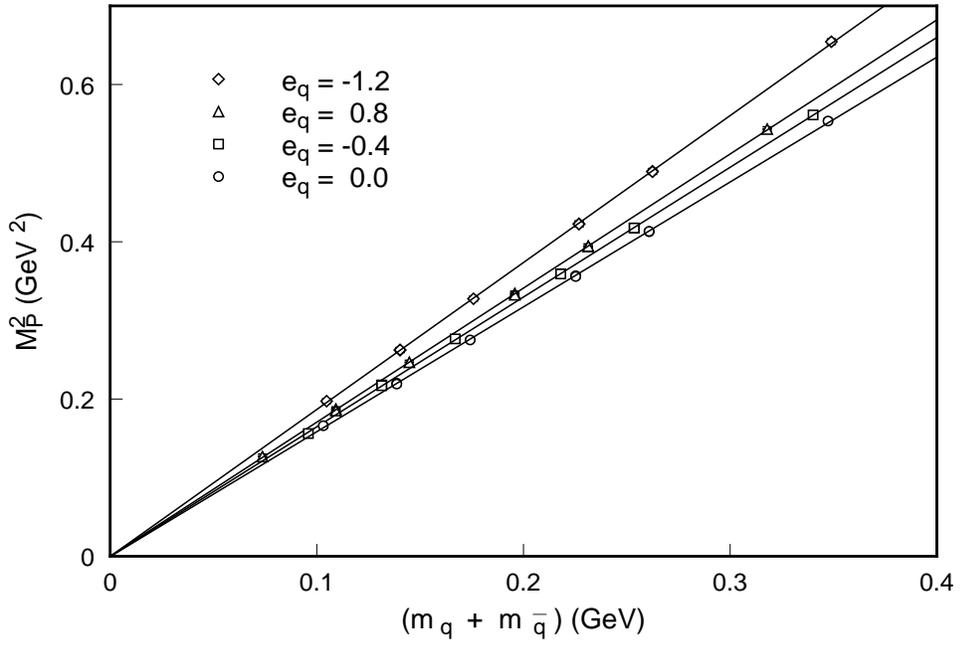,
width=0.95\hsize}$$
\caption{The mass squared, $M_P^2$, (in ${\rm GeV}^2$) for neutral pseudoscalar meson 
versus lattice bare quark masses $m_q + m_{\bar q}$ (in GeV) is shown for 
various quark charges $e_q = 0.0, -0.4, 0.8$ and $-1.2$.}
\label{fig:pi0}
\end{figure}
For the physical values of the quark charges, we expect that an expansion
of the coefficients $A$ and $B$ in (\ref{eq:ChPT})
to first order in $e^2$ should be quite accurate. For the larger values
of QED coupling that we use in our numerical investigation, the accuracy of first
order perturbation theory is less clear: in fact, 
a good fit to all our data
requires small but nonzero terms of order $e^4$, corresponding to 
two-photon diagrams. Comparison of the order $e^4$ terms with those of 
order $e^2$ provides a quantitative
check on the accuracy of QED perturbation theory. 
We have tried including all possible  $e^4$ terms
in the fit, but only retained those which significantly reduce the 
$\chi^2$ per degree of freedom.

According to a theorem of Dashen
\cite{Dashen}, in the limit of vanishing quark mass, the value of $m_P^2$ is
proportional to the square of the total charge.
Thus, we have also allowed the values of the critical hopping parameters for
each of the quark charges to be fit parameters, requiring 
that the mass of the neutral mesons vanish in the chiral limit. 
Thus A takes the form 
$A^{(1)}(e_q+e_{\bar{q}})^2$ to order $e^2$.
(Order $e^4$ terms were not found unnecessary to fit the data.)
The coefficient $B$ in (\ref{eq:ChPT}) which parametrizes the slope of 
$m_P^2$ may also be expanded in perturbation theory. 
Of the five possible $e^4$ terms in $B^{(2)}(e_q,e_{\bar q})$, only the 
$e_q^4, e_q^3 e_{\bar q}$ and $e_q^2 e_{\bar q}^2$ terms
were found to improve the $\chi^2$. The coefficients in A and B,
along with the four values of $\kappa_c$ for the four quark charges, 
constitute a 12-parameter fit to the meson mass values. 

Before discussing the numerical results, we briefly 
describe the formulation of lattice QED
which we have employed in these calculations.
The gauge group in this case is abelian, and one has the choice of
either a compact or noncompact formulation for the abelian gauge action.
Lattice gauge invariance still requires a compact gauge-fermion coupling, but
we are at liberty to employ a noncompact form of the pure photon action
$S_{\rm em}$. Then the theory is  free  in the absence of
fermions, and is always in the nonconfining, massless
phase. An important aspect of  a noncompact formalism  is the necessity for a gauge choice. We use QCD lattice configurations which  have all been converted
to Coulomb gauge for previous studies of heavy-light mesons. Coulomb 
gauge turns out to be both practically
and conceptually convenient in the QED sector as well.

For the electromagnetic action, we take
\begin{equation}
\label{eq:qedact}
S_{\rm em}=\frac{1}{4e^2}\sum_{n\mu\nu}(\nabla_{\mu}A_{n\nu}-\nabla_{\nu}
A_{n\mu})^{2}
\end{equation}
with $e$ the bare electric coupling, $n$ specifies a lattice site, 
$\nabla_{\mu}$ the discrete lattice right-gradient in the $\mu$
direction and 
$A_{n\mu}$ takes on values between $-\infty$ and $+\infty$. Electromagnetic
configurations were generated using (\ref{eq:qedact}) as a Boltzmann weight,
subject to the linear Coulomb constraint
\begin{equation}
\label{eq:couldef}
\bar{\nabla}_{i}A_{ni}=0
\end{equation}
with $\bar{\nabla}$ a lattice left-gradient operator. 
The action is  Gaussian-distributed so it is a trivial matter to 
generate a completely independent set in momentum space,  recovering the real space
Coulomb-gauge configuration
by Fast Fourier transform.  
We fixed the global gauge freedom remaining after
the condition (\ref{eq:couldef}) is imposed 
by setting the $p=0$ mode equal to zero for the transverse modes, and the
$\vec{p}=0$ mode to  zero for the Coulomb modes on each time-slice. (This implies
a specific treatment of finite volume effects which will be discussed below).
The resulting Coulomb gauge field
$A_{n\mu}$ is then promoted to a compact link 
variable $U^{\rm em}_{n\mu}=e^{\pm iqA_{n\mu}}$
coupled to the quark field in order to describe a quark of 
electric charge $\pm qe$.
Quark propagators are then computed for propagation through the combined
SU(3)$\times$U(1) gauge field.

Next we discuss the evaluation of critical hopping parameters for nonzero quark charge.
The self energy shift induced by electromagnetic tadpole graphs 
may be computed perturbatively. The one-loop tadpole graph is (for
Wilson parameter $r$=1 and at zero momentum in Coulomb gauge)
\begin{equation}
\delta m_{EM} = \frac{e^2}{L^4}\sum_{k\neq 0} 
  \{ \frac{1}{4\sum_{\mu}\hat k_{\mu}^2} 
   + \frac{1}{8\sum_i \hat k_i^2} \}
\end{equation}
where $k_{\mu}$ are the discrete lattice momentum components for a 
$L^4$ lattice and $\hat k_{\mu} = \sin(k_{\mu}/2)$. 
This is entirely analogous to the well known QCD  term $\delta m_{QCD}$ 
 \cite{tadpole}.
The mass shift is then given by the sum over multiple insertions 
at the same point, which exponentiates the one-loop graph. 
The usual strong QCD corrections at $\beta = 5.7$ are given in 
this approximation by an overall multiplicative factor of 
$1/(8\kappa_c^{e=0})$.
Together this produces a shift of the
critical inverse hopping parameter of
\begin{equation}
\label{eq:dkap}
\Delta m_c \equiv \left(\frac{1}{2\kappa_c}-\frac{1}{2\kappa_c^{e=0}}\right) 
= \frac{1}{8\kappa_c^{e=0}}(1 - e^{- \delta m_{EM}})
\end{equation}
The contribution from the conventional one loop radiative correction graph is 
found to be about one third the size of the tadpole. 
In Table 1, our numerical results for $\kappa_c$ and the associated 
$\Delta m_c$ is compared with the results using only the perturbative 
tadpole resummed result for the EM interactions(\ref{eq:dkap}).

\begin{table}
\begin{center}
\caption{Calculated shift of critical mass, $\Delta m_c$ versus tadpole estimate 
for neutral pseudoscalar mesons with various quark charges, $e_q$. 
All masses are in lattice units.}
\begin{tabular}{cccc}
 $e_q$ & $\kappa_c$ & $\delta m_c$ & $\sum$ tadpole \\
\hline
~0.0 & 0.16923(3) & --- & --- \\
-0.4 & 0.17130(2) & 0.289(5) & 0.251  \\
~0.8 & 0.17763(3) & 1.118(5) & 0.942  \\
-1.2 & 0.18541(4) & 2.063(6) & 1.912  \\
\end{tabular}
\end{center}
\end{table}


 For charge zero quarks, propagators were calculated at hopping parameter
0.161, 0.165, and 0.1667, corresponding to bare quark 
masses of 175, 83, and 53 MeV
respectively. The gauge configurations are generated at $\beta = 5.7$, and 
we have taken the lattice spacing to be
$a^{-1} = 1.15$ GeV as determined in Ref. \cite{Aida}.
After shifting by the improved perturbative values listed in Table 1, we select the
same three hopping parameters for the nonzero charge quarks. Because this
shift turns out to be very close to the observed shift of $\kappa_c$, the quark 
masses for nonzero charge are nearly the same as those for zero charge.
For all charge combinations, meson masses were 
extracted by a two-exponential fit to the pseudoscalar
propagator over the time range $t = 3$ to 11. 
Smeared as well as local quark propagator sources were used to improve the  
accuracy of the ground state mesons masses extracted.
Errors on each mass value are obtained by a single-elimination jackknife. 
The resulting data is fitted
to the chiral/QED perturbative formula (\ref{eq:ChPT}) by $\chi^2$ minimization.
The fitted parameters are given in Table 2. Errors were obtained by
performing the fit on each jackknifed subensemble.

Aside from very small corrections of order $(m_d-m_u)^2$, 
the $\pi^+-\pi^0$ mass splitting is of purely electromagnetic origin,
and thus should be directly calculable by our method. Because we have
used the quenched approximation,  $u\bar{u}$ and $d\bar{d}$ mesons
do not mix. The  neutral
pion mass is obtained by averaging the squared masses of the 
$u\bar{u}$ and $d\bar{d}$ states. (In full QCD the $u\bar{u}$ and $d\bar{d}$
mix in such a way that the neutral octet state remains a Goldstone
boson of approximate chiral SU(3)$\times$SU(3).
By averaging the squared masses of
$u\bar{u}$ and $d\bar{d}$ in the quenched calculation, we respect 
the chiral symmetry expected from the full theory. By contrast, linear
averaging of the masses would give a $\pi^0$ mass squared nonanalytic
in the quark masses). Thus, to zeroth order in $e^2$, the terms
proportional to quark mass \cite{Weinberg} cancel in the difference
$m_{\pi^+}^2-m_{\pi^0}^2$. This difference is then given quite accurately
by the single term  
\begin{equation}
m_{\pi^+}^2-m_{\pi^0}^2 \approx A^{(1)}e^2
\end{equation}
Using the coefficients listed in Table 2, and the
experimental values of the $\pi^0, K^0,$ and $K^+$ masses,
we may directly solve the resulting three equations for the up,
down, and strange masses. The $\pi^+-\pi^0$ splitting may then be
calculated, including the very small contributions from the 
order $e^2m_q$ terms. We obtain 
\begin{equation}
m_{\pi^+}-m_{\pi^0}= 4.9\pm 0.3 {\rm MeV}
\end{equation}
compared to the experimental value of $4.6$ MeV. (The electromagnetic
contribution to this splitting is estimated \cite{Gasser} to be $4.43\pm0.03$
MeV.)
Our calculation can be compared to the value $4.4$ MeV (for $\Lambda_{\rm QCD}
= 0.3 $ GeV and $m_s = 120$ MeV) obtained
by Bardeen, Bijnens and Gerard\cite{1overN} using large N methods.
The values obtained for the bare quark masses are
\begin{equation}
m_u = 3.86(3),\;\;\; m_d= 7.54(5),\;\;\; m_s = 147(1)
\end{equation}
The errors quoted are statistical only, and are computed by a standard
jackknife procedure. The {\em extremely} small statistical errors reflect the
accuracy of the pseudoscalar mass determinations, and should facilitate
the future study of systematic errors (primarily finite volume,
continuum extrapolation\cite{spacing} and quark loop effects), which are expected to be 
considerably larger. 
The relationship between lattice bare quark masses and the familiar 
current quark masses in the $\overline{MS}$ continuum regularization 
is perturbatively calculable\cite{pert_dm}. 

The presence of massless, unconfined degrees of
freedom implies that the finite
volume  effects in the presence of electromagnetism may be much
larger than for pure QCD. In fact, the corrections are expected to
fall as inverse powers of
the lattice size, instead of exponentially.
We have estimated the size of the finite volume correction phenomenologically
by considering the discussion of Bardeen, et.al\cite{1overN}, which
models the low-$q^2$ contribution to the $\pi^+-\pi^0$ splitting
in terms of $\pi, \rho $, and $A1$ intermediate states. This gives
the splitting as an integral,
\begin{equation}
\label{eq:Bardeen}
\delta m_{\pi}^2 = \frac{3e^2}{16\pi^2}\int_0^{M^2}
                  \frac{m_A^2 m_{\rho}^2}{(q^2+m_{\rho}^2)(q^2+m_A^2)}dq^2
\end{equation}
If the upper limit $M^2$ is taken to infinity, this reproduces the
result of Ref.\cite{Das}, which gives $\delta m_{\pi}=5.1 MeV$.
Even better agreement with experiment is obtained by
matching the low-$q^2$ behavior with the large-$q^2$ behavior
from large N perturbative QCD\cite{1overN}. Here we only use the
expression to estimate the finite volume correction, for which
the low-$q^2$ expression above should be adequate.
To estimate the finite volume effect, we cast this expression as
a four-dimensional integral over $d^4q$ and
then construct the finite volume version of it by
replacing the integrals with discrete sums (excluding
the $q=0$ mode).
For a $12^3\times 24$ box with $a^{-1}=1.15$ GeV, we find
that the infinite volume value of $5.1$ MeV is changed to 
$\delta m_{\pi}=4.8$ MeV, indicating that the result
we have obtained in our lattice calculation should be corrected
upward by about $0.3$ MeV, or about 6\%. In further numerical studies,
we will be able to determine the accuracy of this estimate directly
by calculations on larger box sizes. A study of other systematics
such as finite lattice spacing effects is also in progress, 
and will be reported in a subsequent publication.

For comparison with other results,\cite{Weinberg,leutrev,Donoghue}  
we quote the following mass ratios, which are independent of renormalization 
prescription,
\begin{equation}
\frac{m_d-m_u}{m_s}=.0249(3)\;,\;\;\;\frac{m_u}{m_d}=.512(6)
\end{equation}
With the errors shown, which are statistical only, 
these results differ significantly from the 
lowest order estimate\cite{Weinberg} which uses Dashen's theorem to estimate
the electromagnetic contribution to the kaon splitting to zeroth order.
This lowest order estimate neglects the quark
mass dependence of the electromagnetic
terms, which we have determined by our procedure. Specifically,
the important corrections to the lowest order result come from terms
involving the strange quark mass times the difference of up and down
quark charges. These corrections are determined by the second and third 
terms in $B^{(1)}$ in Table 2.  The Weinberg analysis
predicts that the 4.0 MeV kaon splitting consists of 5.3 MeV from the up-down
mass difference and -1.3 MeV from EM. In our results, the
up-down mass difference contributes 5.9 MeV, with -1.9 MeV from EM. This goes
in the direction indicated by the $\eta\rightarrow 3\pi$ 
decay rate \cite{Donoghue}, although our results do not
deviate as much from the lowest order analysis as those 
of Ref. \cite{Donoghue}, where the quark mass contribution to the 
kaon splitting is estimated to be 7.0 MeV.

\begin{table}
\begin{center}
\caption[tbl:ChPT]{Coefficients of fitting function, Eq.(\ref{eq:ChPT}).
 Terms of order $e_qe_{\bar{q}}^3$ and $e_{\bar{q}}^4$ in $B^{(2)}$ and
 $e^4$ in A were consistent with zero and dropped 
from this fit.
 Numerical values are in ${\rm GeV}^2$ and GeV for
A and B terms respectively.}
\begin{tabular}{cccc}
 Parameter & Fit \\
\hline
 $A  $ & $ 0.0143(10) (e_q + e_{\bar q})^2 $ \\ 
 $B^{(0)}  $ & $ 1.594(11) $ \\
 $B^{(1)}  $ & $ 0.205(22) e_q^2 + 0.071(9) e_q e_{\bar q} 
+ 0.050(7) e_{\bar q}^2 $ \\
 $B^{(2)}  $ & $ 0.064(17) e_q^4 + 0.033(6) e_q^3 e_{\bar q} 
- 0.031(4) e_q^2 e_{\bar q}^2 $ \\
\end{tabular}
\end{center}
\end{table} 

In the present work we have focused on the pseudoscalar meson masses. This 
is the most precise way of determining the quark masses as well as 
providing an important test of the method in  the $\pi^+-\pi^0$ splitting.
Further calculations of electromagnetic splittings in the vector mesons and
the baryons, as well as in heavy-light systems, are possible
using the present method. This will provide an extensive opportunity
to test the precision of the method and gain confidence in the results.
Further study of electromagnetic properties of hadrons in lattice QCD,
such as magnetic moments and form factors, is also anticipated.

We thank Tao Han, George Hockney, Paul Mackenzie and
Tetsuya Onogi for contributions to our effort. 
AD was supported in part by the National Science Foundation
under Grant No. PHY-93-22114.
HBT was supported in part by
the Department of Energy under Grant No.~DE-AS05-89ER 40518.
This work was performed using the ACPMAPS computer
at the Fermi National Accelerator Laboratory, which is operated by
Universities Research Association, Inc., under contract DE-AC02-76CHO3000.



\end{document}